\documentclass[11pt]{article}
\usepackage{graphicx}
\usepackage{psfrag}
\def\at{\string@}
\def\t2#1{\underline{\underline{#1}}}
\newcommand{\be}{\begin{equation}}
\newcommand{\ee}{\end{equation}}
\newcommand{\fff}{{\mathcal F}}
\newcommand{\ggg}{{\mathcal G}}
\newcommand{\hhh}{{\mathcal H}}
\newcommand{\ddd}{{\mathcal D}}
\makeatletter
\renewcommand{\section}{\@startsection {section}{1}{1cm}{-2.5ex \@plus -1ex
\@minus -.2ex}{2.1ex \@plus .2ex}{\reset@font\bfseries }}
\renewcommand{\fnum@figure}{\small\textbf{Figure~\thefigure}}
\renewcommand{\@biblabel}[1]{\quad #1.}
\renewcommand{\author}[1]{
\begin{center}
{\bf #1}
\end{center}
\par
}
\newcommand{\address}[1]{%
\begin{center}
#1
\end{center}
}

\renewcommand{\title}[1]{\begin{center}
    {\large{\bf #1}}
     \end{center}
\medskip
} 

\renewcommand{\abstract}[1]{%
\begin{center}%
\begin{minipage}{16cm}%
  {\small%
   \parindent=0pt%
    {\bf ABSTRACT: }%
     #1\par}%
\end{minipage}%
\end{center}%
}%
\newcommand{\resu}[1]{%
\begin{center}%
\begin{minipage}{16cm}%
  {\small%
   \parindent=0pt%
    {\bf R\'ESUM\'E : }%
     #1\par}%
\end{minipage}%
\end{center}%
}%
\def\motscles#1{
\begin{center}
{\begin{minipage}{16cm}
                {\small%
                 \parindent=0pt
                 {\bf MOTS-CL\'ES : }
                 #1\par
                }
\end{minipage}
}
\end{center}
\vskip 2em \par
}
\def\keywords#1{
\begin{center}
{\begin{minipage}{16cm}
                {\small%
                 \parindent=0pt
                 {\bf KEYWORDS: }
                 #1\par
                }
\end{minipage}
}
\end{center}
\vskip 2em \par
}
\makeatother
\def\be{\begin{equation}}
\def\ee{\end{equation}}

\begin{document}
\title{ QUASISTATIC RHEOLOGY AND THE ORIGINS OF STRAIN.}
\author{Jean-No\"el ROUX and Ga\"el COMBE}
\address{Laboratoire des Mat\'eriaux et des Structures du G\'enie Civil\\
(Unit\'e mixte LCPC-ENPC-CNRS, UMR113)\\
2 all\'ee Kepler, cit\'e Descartes\\
77420 Champs-sur-Marne, France
}
\resu{
On confronte l'approche microscopique par simulation num\'erique discr\`ete
des mat\'eriaux granulaires de type solide \`a leurs propri\'et\'es rh\'eologiques
macroscopiques. On cite des syst\`emes mod\`eles dont les r\'eponses, en
d\'eformation, à un incr\'ement de contrainte diff\`erent qualitativement,
bien que la r\'epartition des efforts à l'\'equilibre soit tr\`es similaire.
Des r\'esultats sur la sensibilit\'e aux perturbations
des r\'eseaux de contact \'elastoplastiques
permettent de distinguer
deux r\'egimes rh\'eologiques, selon que leurs intervalles de stabilit\'e, en termes de
contraintes, se r\'eduisent ou non à z\'ero  dans la limite thermodynamique
(`fragilit\'e' macroscopique). On en \'evoque de possibles cons\'equences.
}
\abstract{Features of rheological laws applied to solid-like
granular materials are recalled and confronted
to microscopic approaches via discrete numerical simulations.
We give examples of model systems with very
similar equilibrium stress transport properties -- the much-studied force
chains and force distribution --
but qualitatively different strain responses to stress increments.
Results on the stability of elastoplastic
contact networks lead to the definition of two different rheological
regimes, according to whether a macroscopic
fragility property (propensity to rearrange under arbitrary small
stress increments in the thermodynamic limit)
applies. Possible consequences are discussed.
}
\motscles{D\'eformation, loi de comportement, simulations num\'eriques}
\section{Scope}
This is a brief introduction to the
rheology of solid-like granular materials in the quasistatic regime, with a
special emphasis on the microscopic origins of strain, and on discrete
numerical simulations of model systems.
Rather
sophisticated macroscopic phenomenological laws have been proposed~\cite{GDV84,Darve87,DMWood}, but, in spite
of many microscopic studies~\cite{KI01,HHL98}, with numerical tools~\cite{CJ01} in particular, their
relation to grain-level physical phenomena is not fully understood.
Consequently, we mostly address basic, qualitative aspects on model systems. Moreover, we specialize on
cohesionless, nearly rigid grains, and to small or moderate strain levels (excluding
continuous, unbounded plastic flow).
Despite the many insufficiencies of present-day modelling attempts, interesting
directions for future research, elaborating on preliminary results, can be suggested.
We recall a few basic concepts (section 2), some of the macroscopic phenomenology of
solid-state granular mechanics (part 3),
and the necessary elements of a microscopic model (section 4). Then, properties of
simple model systems studied by numerical means, in the large system limit, are discussed
both in frictionless (section 5) and in frictional (section 6) systems. Section 7 suggests
broader perspectives and speculations.
\section{The constitutive law approach: basic ideas.}
On setting out to identify a constitutive law for a solid material, one has to rely
on some postulates that are worth recalling in the context of granular materials.
Such a law should \emph{locally} relate stresses $\t2{\sigma}$ to strains $\t2{\epsilon}$ ,
or, more appropriately for granular systems, stress rates
$\t2{\dot \sigma}$ and strain rates $\t2{\dot \epsilon}$ should
determine each other for a given internal state of the system. This state is to be
conceived of as specified once the values of some state variables (a finite number $p$ of
quantities, including the stress tensor itself, that
exhaust the macroscopic description of the system) are known. One may write, at each time
$t$:
\be
\t2{\dot \sigma}({\bf x},t) = \fff \left( \t2{\dot \epsilon}({\bf x},t),\t2{\sigma}({\bf x},t),
\left\{\alpha({\bf x},t)\right\}\right)
\mbox{\ \ or \ \ }
\t2{\dot \epsilon}({\bf x},t) = \ggg \left( \t2{\dot \sigma}({\bf x},t),\t2{\sigma}({\bf x},t),
\left\{\alpha({\bf x},t)\right\}\right)
\label{eqn:loiconstgene}
\ee
${\bf x}$ standing for any `point', in the sense of continuum mechanics, in the sample,
i.e. a representative volume element from the microscopic point of view. $\{\alpha\}$ is the set
of unspecified state variables $(\alpha _i)_{1\le i\le p}$. Their evolution should
be ruled by similar equations:
\be
\dot \alpha _i({\bf x},t) = \hhh \left( \t2{\dot \epsilon}({\bf x},t),\t2{\sigma}({\bf x},t),
\left\{\alpha({\bf x},t)\right\}\right).
\label{eqn:loiconstgene2}
\ee
Once eqns.~\ref{eqn:loiconstgene} and~\ref{eqn:loiconstgene2} are given, and supplemented with the appropriate
boundary conditions,
one also needs the initial values of
$\t2{\sigma}$ and $\alpha$ to be able to predict the evolution of the system for, say, a prescribed history
of stress. The prediction of the initial state of the system is in general beyond the scope of the rheological
laws we are dealing with, as it is the result of a process that might involve rapid flow. (One exception is the
construction of a sample under gravity by successive deposition of thin layers at the free surface. If the initial
state of a freshly deposited layer is known, one may apply solid-like rheology to the rest of the sample, which
deforms very little under the weight of the new layer, 
and thus, iteratively solving the appropriate boundary value problem,
calculate the initial state of a whole system as a result of its construction
history. This procedure can be applied to silos~\cite{Ragneau} and granular piles~\cite{Boufellouh}).

The suggestions, put forward in the recent literature~\cite{CWBC98}, to 
look for direct relationships between stress components that result from the construction history of a sample,
are attempts
to model the assembling procedure, rather than the
response to stress increments. The proposed relations are not constitutive laws in the sense
of eqns.~\ref{eqn:loiconstgene} and~\ref{eqn:loiconstgene2}: they 
ignore strains, and they are not local (they depend on sample shape and boundary conditions). 

One thus needs some a-priori knowledge of the initial stresses:
however small the components of the stress tensor,
the orientation of its principal axes and its level of anisotropy
are important. Cohesionless grains do not spontaneously assemble in any `natural state', once
submitted to some externally imposed stresses they form packings the structure of which depends on those stresses.
Functions $\fff$, $\ggg$ and $\hhh$ of relations~\ref{eqn:loiconstgene} and~\ref{eqn:loiconstgene2}
must be discontinuous at $\t2{\sigma}=0$.
In rheometric experiments one needs in principle to check for sample homogeneity.
\section{Macroscopic aspects}
\begin{figure}
 \centering 
\psfrag{v}[c]{$\dot \epsilon_1$, $\sigma_1$}
 \psfrag{a}[c]{$\sigma_3$}
 \includegraphics[width=5cm]{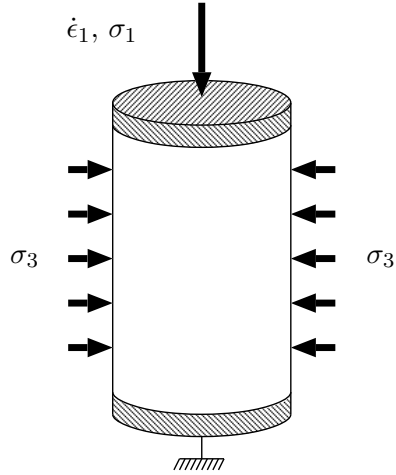}
\caption{Sketch of a triaxial experiment.
 \label{fig:triax1}
}
\end{figure}

\begin{figure}
 \centering 
 \psfrag{x}[l]{$\epsilon_1$}
 \psfrag{y1}[l]{$q = \sigma_1 - \sigma_3$}
 \psfrag{y2}[l]{$-\epsilon_v$}
 \psfrag{p}[l]{\footnotesize `peak'}
 \includegraphics[width=5.5cm]{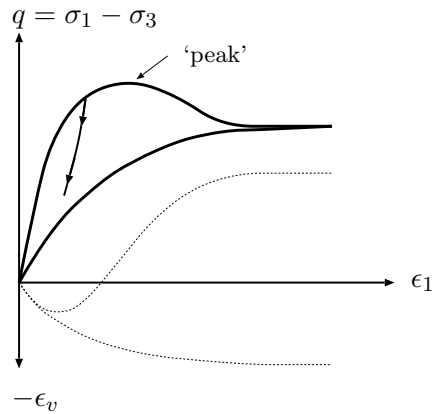}
\caption{Schematic variations of deviator $q$ (continuous curves) and volumetric strain $\epsilon _v$ 
(dotted lines) for a dense and a loose sand. The curve marked with arrows
is observed on decreasing $q$ or $\epsilon _1$.
 \label{fig:triax2}           
}
\end{figure}

Constitutive laws like eqns.~\ref{eqn:loiconstgene}
and~\ref{eqn:loiconstgene2} are studied in soil mechanics~\cite{DMWood,HHL98,BH94}.
In order to extract some information on such laws from experiments, it is convenient to choose configurations
in which stresses and strains are expected to be homogeneous. This leads to the design of rheometers, the
most often employed one in soil mechanics being the triaxial apparatus, sketched on fig.~\ref{fig:triax1}.
Samples are submitted to axisymmetric states of stress,
the axial stress $\sigma_1$, or the axial strain $\epsilon_1$,
are controlled via the relative motion of the end platens,
while the lateral pressure $p=\sigma_2 = \sigma_3$ is exerted
through a flexible membrane by a fluid. With some care
(e.g., measuring strains directly on the sample in the
central part away from the rigid platens) 
it is possible, with the most sophisticated devices, to record strains
with an accuracy of the order of $10^{-6}$~\cite{dBGSC99,HI96}.
Other rheometers~\cite{LA87} are the `true triaxial' apparatus, which
may impose three different principal stresses
to a cubic sample, and the `hollow cylinder' apparatus, which allows
for rotation of the principal axes of stress and
strain.
In a typical triaxial experiment, one starts from a given state with, e.g., hydrostatic stress
($\t2{\sigma}=p\t2{{\bf 1}}$). Then, most often, $\epsilon_1$
is increased at a constant (slow) rate, while lateral
pressure $p$ is maintained constant. Axial stress $\sigma_1$ -- or, equivalently, deviator $q=\sigma_1-p$ -- and
the lateral strain $\epsilon_2=\epsilon_3$ (or, equivalently, the relative volume increase,
$\epsilon_v=-tr(\t2{\epsilon})$)
are measured. Evolutions of $q$ and $\epsilon_v$ as $\epsilon _1$ monotonically
increases are schematically represented
on fig.~\ref{fig:triax2}. 
While the density and the deviator steadily increase, in a loose sample,
until asymptotic constant values are reached, 
dense samples are initially contractant, then dilatant and the deviator curve passes through a maximum. 
From the beginning, the increase of $q$ with $\epsilon _1$ is not reversible:
if the direction of deformation is reversed,
the same curve is not retraced back: the decrase of the deviator is steeper,
with a slope comparable to that of the
tangent at the origin of coordinates on fig~\ref{fig:triax2}. 

As $\epsilon _1$ increase, the curves approach a plateau, corresponding to a final attractor that 
is called the `critical state'~\cite{DMWood} and deemed independent of
the initial conditions (density and deviator should coincide for the loose and the dense sample
on fig.~\ref{fig:triax2} at large $\epsilon _1$).
However, the approach to this critical state is often hidden,
in dense samples at least, by instabilities
leading to strain localization in `shear bands' (whose thickness is of the order of a few grain diameters).
The development
of these localizations was observed by X-ray tomography~\cite{DCMM96}. It is sensitive to sample shape and 
boundary conditions, but usually occurs in the
vicinity of the observed `stress peak'. Localization in loose samples,
if it exists, is far less conspicuous.
As sample homogeneity is lost, localization precludes the interpretation of
rheometric tests in terms of constitutive laws,
which should therefore be restricted to the `pre-peak' part of the curve. 

An important feature of rheological tests on solid-like granular samples is their
independence on physical time (see however, the remarks of section 7). 
Replacing $t$ by $\varphi(t)$ with a monotonically increasing function $\varphi$ does not
change stress-strain curves. Functions $\fff$, $\ggg$ , $\hhh$ should therefore be homogeneous of degree one in 
$\t2{\dot \epsilon}$ or in $\t2{\dot\sigma}$. Such tests are supposed to be \emph{quasi-static}, as a sequence of
equilibrium states is explored. Dynamical characteristics, such as masses, are regarded as irrelevant. Dry
sands and water-saturated ones, provided static properties are the same, should exhibit the same behaviour.

Eqns.~\ref{eqn:loiconstgene} and~\ref{eqn:loiconstgene2}, with due account for
material symmetries, remain extremely general. The nature of internal variables
$\alpha$ is not easy to guess if only stresses and strains are measured. 
The packing fraction is often chosen, because of its influence on the behaviour (fig.~\ref{fig:triax2}).

To account for irreversibilies and failure, the most commonly invoked laws are of the elastoplastic
family. Those (although usually written differently) can be cast in the form of
eqns.~\ref{eqn:loiconstgene} and~\ref{eqn:loiconstgene2}. We do not review elastoplasticity here. Its
application to soil mechanics is presented in ref.~\cite{DMWood}. Connections
to limit analysis (calculation of a limit load beyond which unlimited plastic flow
occurs) are discussed in~\cite{SAL77}. Ref.~\cite{VER98} is a pedagogical introduction
with examples of calculations with simplified laws. The complex behaviour sketched on fig.~\ref{fig:triax2}
requires quite sophisticated elastoplastic laws, with many parameters. Those laws should be
\emph{non-associated}~\cite{VER98}, and involve \emph{work hardening}.
Very roughly speaking, this means that the direction of 
plastic irreversible strains is not simply related to the failure condition on stresses,
and that failure is a gradual process.
A simplified law with correct qualitative properties in terms of global failure under monotonically varying
loads~\cite{VER98} involves 
4 parameters. A more elaborate and quantitative one, Nova's law~\cite{NO82}, requires 8 parameters.
Incremental~\cite{GDV84} or hypoplastic~\cite{WBK96} laws are
less traditional. They
directly state relations like eqns.~\ref{eqn:loiconstgene}
and~\ref{eqn:loiconstgene2}. $\fff$ and $\ggg$, although positively homogeneous of degree one,
should be non-linear functions of $\t2{\dot \epsilon}$ and $\t2{\dot \sigma}$ respectively, to account for
irreversibility. This is directly postulated in such approaches, while elastoplastic laws describe
such a behaviour through work hardening. Ref.~\cite{WBK96} defines a simplified hypoplastic law with 5 parameters.

Obviously, it is highly desirable to identify parameters with a physical meaning,
connected with the microscopic mechanisms of
deformation under stress. This would ease and guide the choice of a constitutive law, contribute
to assess its range of validity (e.g., in terms of stress magnitudes), and reveal the influence of microscopic
characteristics of a given material on its macroscopic behaviour.
\section{Ingredients of a microscopic model, discrete simulations.}
Numerical simulation methods are described in ref.~\cite{CJ01}. They deal with
simple models of granular materials, suitable to investigate the microscopic
origins of constitutive laws. Here, we mainly discuss the simple cases of discs (2D)
or spheres (3D). 
Three different kinds of ingredients are needed: geometric, static and dynamical ones.
First, grain shape and polydispersity have to be specified. 
Then, static parameters are those defining equilibrium contact laws.
In the most simple model, a normal stiffness constant $K_N$ expresses
a proportionality between normal force $f_N$
and normal deflection $h$ of a contact (which is modelled as an interpenetration depth),
a tangential stiffness $K_T$ incrementally relates tangential contact forces $f_T$
to tangential relative displacements,
and Coulomb's condition $\vert f_T \vert \le \mu   f_N$ should be satisfied, with a friction
coefficient $\mu$, sliding being allowed (such that the work of $f_T$ is negative) if it holds as an equality. 
One may regard stiffness constants as a mere computational trick to forbid grain interpenetration. 
They may also be chosen with correct order of magnitudes, comparing them
to typical estimated values of $df_N/dh$ in contacts, under some given stress level. It can even be attempted to
implement accurate contact laws, such as the Hertz-Mindlin-Deresiewicz~\cite{TY91}
ones for smooth elastic spheres with friction.

Finally dynamical parameters are related to inertia (masses, moments of inertia) and kinetic energy dissipation
(e. g. viscous damping in contacts). In
slow, quasi-static evolutions, those parameters should be irrelevant, the behaviour should be determined
by the geometric data, along with $\mu$, ratio $K_T/K_N$ and parameter $K_N/P$ (in 2D) or $K_N/(Pd)$ (in 3D), which
measures the deflections of contacts relative to grain diameter $d$ under typical forces~\cite{Gael}.  

The most widely used simulation methods~\cite{CJ01}, molecular dynamics (MD), or contact dynamics,
rely on time integration of dynamical
equations of motion. They have been used to simulate biaxial (in 2D)~\cite{LJ00} or triaxial~\cite{TH00} tests. 
Such calculations are usually made at constant axial strain rate, assuming the system remains close enough to
equilibrium at any time for the evolution to be regarded as quasi-static.
They cope with a few hundreds or a few thousands
of grains. They successfully produce stress-strain curves whose broad features,
on the scale of $\epsilon \sim 10\%$,
are those of fig.~\ref{fig:triax2}. For instance Thornton's simulations~\cite{TH00}
yield a maximum deviator criterion
that coincides with some experimental observations. However, stress-strain curves, for given loading histories, are
still rather noisy on a smaller scale ($\epsilon \sim 1\%$), especially as the peak is approached
(strain values corresponding to the peak deviator are not accurate).
It seems
necessary to study the form of such curves and investigate the regression of fluctuations with greater care,
for two reasons: first, one might wish to obtain accurate estimations of
rheological laws in the macroscopic limit; then
reliable numerical data on the effect of perturbations on equilibrium states would give insight on the microscopic
mechanisms of deformation. Those should be accounted for in theoretical
attempts to relate rheological laws to grain-level
phenomena. The next sections are therefore devoted to biaxial compressions with 2D systems of
disks, in which deviator $q=\sigma _1-\sigma _2$ (keeping the notations introduced for the triaxial test) 
is stepwise increased and one studies the effect of small stress increments imposed on
equilibrium configurations.
\section{Response to stress increments: frictionless grains}
Assemblies of frictionless grains are particularly appealing because of two remarkable properties,
that are established and discussed in ref.~\cite{JNR2000}. The first one, the {\em absence
of hyperstaticity} in the limit of large contact stiffness, means that the contact network is barely sufficient to
support stresses, and that the sole condition that only closed contacts can transmit a force, along with
the force (and torque) balance equations, is sufficient to calculate all contact forces. The second (the standard
mechanical energy minimization property) states, for rigid grains, that {\em the potential energy
of external forces has to be minimized} under the constraints of
no interpenetration. It allows to discuss the stability of equilibrium states.

\begin{figure}
 \centering 
 \includegraphics[width=5cm]{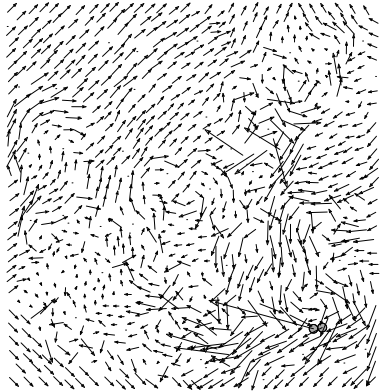}
\caption{Velocity field at the beginning of a rearrangement in a system of 1024 disks in a square box
undergoing biaxial compression. The contact between the disks marked in gray (bottom right) is opening.
 \label{fig:vitesses}
}
\end{figure}

\emph{Together}, both properties entail~\cite{JNR2000} that force-carrying
structures in assemblies of rigid frictionless and
cohesionless disks (in 2D) or spheres (in 3D) are isostatic: there is a one to one correspondence between
external forces and contact forces, and between relative normal velocities in the contacts and grain velocities.
We exploited this~\cite{CR2000} to study the response of disordered systems
of disks to stress increments, by a purely
geometric procedure we called the `geometric quasi-static method' (GQSM).
In such systems, the stress-strain curve is a
staircase and the procedure tracks the elementary steps. In stability intervals,
the rigid contact structure supports
the stress without motion, until one contact force becomes negative. This contact has then to open,
initiating a rearrangement, hence a strain increment. Motion stops when another contact closes and a new stable
equilibrium is reached. In fact, this might require several contact replacements, which are operated
one by one in this algorithm. On opening one contact, the ensuing velocity field is, up to a positive factor,
geometrically determined. One example is displayed on fig.~\ref{fig:vitesses}.
Those fields form complex vortex patterns extending through the whole sample. Displacement fields corresponding to
small strain increments have the same aspect. 

Statistics of stress and strain steps for the beginning of a biaxial compression
(close to the isotropic state of stress) were studied~\cite{CR2000,Gael}. It was found that 
intervals of stability $\delta q$ are exponentially distributed and scale with
the number of disks $N$ as $N^{-\alpha}$ with $\alpha\simeq 1.1$,
while `axial' (conjugate to the largest principal stress
that is being incremented) strain steps are power-law distributed, the density function decrasing as 
$(\delta \epsilon)^{-(1+\mu)}$, with $\mu\simeq 0.5$ for large values,
and scale as $N^{-\beta}$ with $\beta\simeq 2.1$.

As stability regions in stress space dwindle to nothing in the thermodynamic limit
(a macroscopic `fragility' property~\cite{CWBC98b,JNR2000}), equilibrium states will be
rather elusive: it is impossible, in a real experiment, to control stress levels with perfect accuracy.
Although
each equilibrium configuration is rigid, any level of noise in a
macroscopic system should generate fluctuations, because
rigid configurations become unstable, and the system should keep visiting several equilibrium states.

\begin{figure}
 \centering 
 \includegraphics[width=7cm]{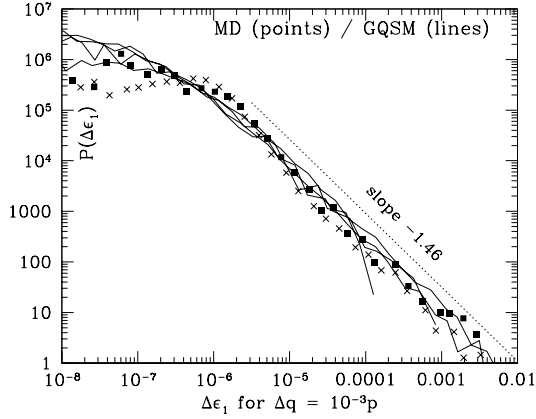}
\caption{Distribution of axial strain increments corresponding to fixed $\Delta q =10^{-3}p$ measured
for 4 sample sizes (up to $N=4900$) with GQSM and 2 sizes (1000 and 3000) with MD. Note the independence
on the numerical method (for large values) and the absence of convergence, as $N$ increases, 
to a deterministic response.
 \label{fig:levy}
}
\end{figure}

Since the power law distribution of strain increments does not admit a mean value, the accumulation of successive
$\delta \epsilon$ steps generates a L\'evy process~\cite{BG90},
and the strain step corresponding to a given deviator
increment remains impredictable. Although, as $N$ increases, the typical size of steps
decreases, the staircase stress-strain curve does not become smooth because of the statistical importance of large
strain increments. It was also observed~\cite{CR2000}
that the statistics of strain variations corresponding to given
stress intervals do not depend on $N$. Moreover, this distribution
is the same for the GQSM and for a more conventional MD calculation
(with nearly rigid grains, $K_N/P=10^5$). As MD results, introducing additional
static and dynamical parameters, are statistically indistinguishable
from GQSM ones, one may conclude that the mechanical response is determined by geometry alone.
These results are summarized on fig.~\ref{fig:levy}

This very singular behaviour -- a constitutive law
cannot be defined -- calls for additional investigation of its origins and range of validity. It is worth pointing
out that the observation of the large $\delta \epsilon$ values and the L\'evy-stable distribution are due to
the rearrangements in which several (sometimes many) contacts have to be replaced.
If we now adopt the approximation of
small displacements (ASD) in which displacements away from a reference configuration are dealt with as
infinitesimal, and hence normal unit vectors between neighbouring disks are kept constant, then it can be
shown~\cite{JNR2000} that only one contact replacement will be enough to reach the next equilibrium state.
Then, the distribution
of strain steps admits a first and a second moment, and the staircase,
within the ASD, should approach a smooth curve.

In fact, the ASD with frictionless grains has deep consequences: it entails~\cite{JNR2000}
the uniqueness of equilibrium states. There
is no need to obtain a stress-strain curve in an incremental way, as stresses and strains are in one-to-one
correspondence. Uniqueness also implies the absence of irreversibility, since there is no history dependence.
Assemblies of slightly polydisperse disks placed on a regular triangular network
were studied within the ASD (with the unperturbed lattice as reference). The approximation
is well controlled in that case because of the small polydispersity parameter.
This system might be called elastic (although disks are rigid). Its behaviour was shown~\cite{JNR97b,JNR2000}
to be analogous to that of a mobile point requested to stay within a convex part $\ddd$, limited by a
smooth surface $\Sigma$,
of a three-dimensional space (the analog of the set of values permitted by
impenetrability conditions in strain space). Once submitted to an external force
(its 3 coordinates are the analogs of the stress components) its equilibrium position is the point of
$\Sigma$ where the tangent plane is orthogonal to the force. Hence a 
smooth correspondence between force and displacements. Upon incrementing the force, the displacement
is inversely proportional to the curvature of $\Sigma$. As a consequence of these
properties~\cite{JNR2000},
the macroscopic response (displacements) to some small localized force superimposed on a pre-imposed
stress field (Green's function) is, for this model, 
the solution to an elliptic boundary value problem (akin
to elastic problems for incompressible materials). Within the ASD, finding an equilibrium state amounts to solving
a convex minimization problem. This is also the case, without the ASD, for networks of cables, for which the same
kind of elasticity applies. 

Let us summarize the main conclusions of these studies on rigid frictionless grains.

(\emph{i}) The stress-strain curve is a staircase with phases of stability, with
just enough contacts to carry the forces, alternating with rearrangements, with just enough contact openings to
allow some deformation. Those rearrangements are non-local events, involving the whole system.
(\emph{ii}) Any macroscopic stress perturbation causes some rearrangement.
(\emph{iii}) No deterministic constitutive law applies to disordered assemblies of rigid, frictionless disks. 
(\emph{iv}) The mechanical response is determined by the sole geometric data. 
(\emph{v}) Different systems might exhibit equilibrium states with extremely similar properties
(in terms of force
distributions) and both satisfy properties \emph{i} and \emph{ii}. However their mechanical behaviour
might be drastically different: conclusion \emph{iii} applies to disks without the ASD, whereas disks within the
ASD or cable networks abide by some form of elasticity, rearrangements being reversible. 
\section{Response to stress increments: grains with friction}
\begin{figure}
 \centering 
 \includegraphics[width=6.5cm]{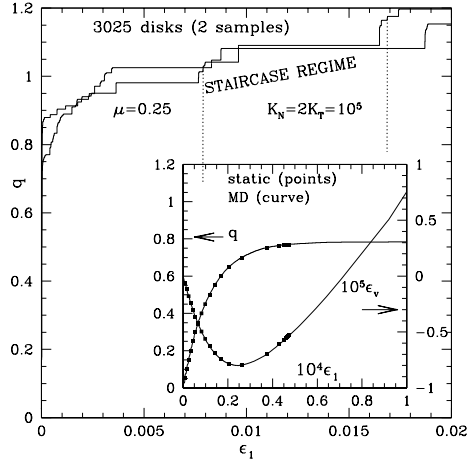}
\caption{Main plot: aspect of deviator/axial strain curves for $\epsilon _1 \le 0.02$ (MD calculations). The dotted
lines are results obtained with the static method on going backwards on the same stress path ($q$ decreases).
The inset
 -- note the blown-up $\epsilon$ scales -- shows $q$ and $\epsilon_v$ versus $\epsilon _1$ in the strictly
quasi-static regime
 \label{fig:dessmarfro}
}
\end{figure}

\begin{figure}
 \centering 
 \includegraphics[width=6.5cm]{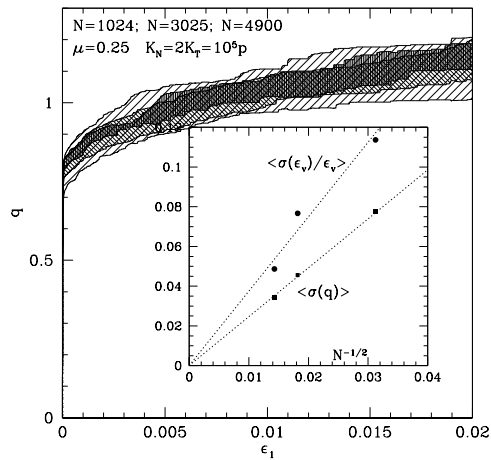}
\caption{$q(\epsilon _1)$ for 3 sample sizes: the larger $N$, the darker is the shaded
zone extending to one standard deviation
above and below the average. The inset displays the average standard deviation
throughout the $\epsilon _1$ range, for $q$ and for the volumetric strain, versus $N^{-1/2}.$
\label{fig:courbstat}
}
\end{figure}

Let us now report some results on disordered systems of nearly rigid ($K_N/P=10^5$) systems of disks
with a friction coefficient $\mu>0$ in the contacts~\cite{CR01,Gael}.
Just like in the frictionless case, one may either try to track elementary
stability intervals and rearrangements, or resort to molecular dynamics (introducing additional parameters).
However,
the specific properties of frictionless systems are lost: the contacts may transmit tangential forces and the
network is hyperstatic; potential energy is no longer minimized at equilibrium. To discuss the stability of a
given contact network, one needs to introduce elasticity in the contacts. One can then perform
static elastoplastic calculations: the system is to be regarded as a network of springs, plastic sliders and
no-tension joints, the displacements and rotations of the disks can be computed for each applied stress
increment, via an iterative process~\cite{Narimane}. 
Such studies of elastoplastic networks~\cite{Narimane,KAK01}
with static methods are surprisingly rare, especially in comparison to the
vast literature on numerical simulations of elastic networks and brittle fracture (see e.g. \cite{CRG90}) 
The systems studied in~\cite{CR01} were prepared without friction, and are thus
very dense. Several simple sizes were studied ($N$ ranging from $1000$ to $5000$), and it was found that the initial
contact network, corresponding to isotropic stresses at the beginning of the biaxial compression, is able to
support a considerable stress deviator ($q/p=0.81\pm 0.06$ for $\mu=0.25$) in the large system limit. Therefore,
such dense systems with friction are not fragile in the sense of section 5. Stress-strain curves for
$\mu=0.25$, $K_T/K_N=0.5$ for the beginning of the biaxial compression are shown on fig.~\ref{fig:dessmarfro}.
In this regime, that we call {\em strictly quasi-static}, the  curve is smooth, the successive equilibrium
configurations form a continuum. The scale of strains is set by the stiffness constants in the contacts. The 
behaviour is inelastic and irreversible from the beginning of the biaxial compression, as the proportion of
sliding or opening contacts steadily increases.
Eventually, some instability occurs, the initially present contacts
can no longer support the stresses. Interestingly, this appears to happen for a deviator value that does not
sensitively depend on stiffness constants~\cite{Gael}.
To proceed further (as the current state
of the static algorithm does not clearly determine one direction of instability and no
analog of the GQSM is available), we resorted to molecular
dynamics. Successive equilibria corresponding to stepwise increasing deviator values were obtained, and a
staircase-like stress-strain curve was observed, signalling the frequent occurrence of instabilities and strain
jumps (fig.~\ref{fig:dessmarfro}). 
To check whether a well-defined stress-strain relation is approached as $N\to\infty$, averages and
mean standard deviations of $q(\epsilon _1)$ for $0\le \epsilon _1\le 0.02$ were computed with
several samples of different sizes, and the
results (fig.~\ref{fig:courbstat}) do indicate that a smooth curve is approached. Similar results are
obtained for $\epsilon _v$ versus $\epsilon _1$. 
Any strictly quasi-static interval appears as a vertical segment on fig.~\ref{fig:courbstat}. The existence
of a limit for large $N$ requires in fact the fragility property to apply in the staircase regime. However, 
it is expected that although any \emph{positive} increment $\delta q$ will entail a rearrangement in a macroscopic
system, contact structures will withstand finite \emph{negative} $\delta q$'s. Fig.~\ref{fig:dessmarfro} shows
that relatively large $q$ intervals could be accessed by static calculations, from
intermediate equilibrium configurations in the staircase regime, upon decreasing $q$. Moreover,
it is observed that many contacts stop sliding on reversing the motion. 
It would be interesting to investigate
the response to differently oriented stress increments, and to delineate the 
(history-dependent) strictly quasi-static, non-fragile domain around a given equilibrium state. 
Taking the mobilization of friction
into account -- i.e., replacing Coulomb's inequality by an equality for all sliding contacts -- it can be
observed that the indeterminacy of forces is greatly reduced within the staircase regime in the monotonic
biaxial compression~\cite{LJ00,CR01}, which suggests an analogy with isostatic frictionless systems. 
Grain motions, rearrangements and spatial distribution of strains were studied by Williams
and Rege~\cite{WR97}, and by Kuhn~\cite{KU99}. Similar patterns as those of fig.~\ref{fig:vitesses} were observed.
Thin non-persistent (unlike shear bands) `microbands' concentrating the strain were also reported~\cite{KU99}. 

There is therefore some evidence that conclusions (\emph{i}), (\emph{ii}), and (\emph{iv}) of section 5 are still
valid in systems with friction \emph{within the fragile `staircase' regime}, the essential differences being the role
of the friction coefficient itself (the behaviour appears to be essentially determined by the geometry \emph{and}
$\mu$), the existence of a well-defined macroscopic stress-strain curve and that of strictly quasi-static
regimes within which a given network of contacts is able to support a finite stress range, all strains being
due to the finite stiffness of the grains themselves.
\section{Some perspectives and speculations.}
The \emph{non-local} aspect of rearranging events, reflecting the strong steric hindrance in dense packings of 
impenetrable bodies, could be expected to preclude the definition of a local law. However, one might think that
a great number of such long-distance correlated motions
of very small amplitudes could, once aggregated, build up a strain field devoid of long-range correlations. 
Specifically, strains could be localized on some `microband' pattern (as reported by Kuhn~\cite{KU99}) during
one elementary rearrangement, but the random superposition of lots of such non-persistent structures could destroy
the long-distance correlations. Similar ideas were followed by T\"or\"ok \emph{et al.}~\cite{TKKR00}: their schematic
model assumes macroscopic deformation to result from an accumulation of slides on temporary slipping surfaces.
It would be interesting to test such a scenario, which requires an accurate numerical computation of directions
of instabilities of granular assemblies with a given contact network. That `deformation consists in a number
of arrested slides' is an idea already put forth by Rowe in 1962 (ref.~\cite{RO62},p. 514). His classical `stress-dilatancy
relation'~\cite{RO62} could thus be founded on a microscopic analysis.

In parallel to the analysis of stress-strains relationships we have been reporting here
(we focussed on eqn.~\ref{eqn:loiconstgene}), microscopic studies
have tried to define internal state variables of granular systems and to relate them to stresses and strains
(i.e. suggesting a form of eqn.~\ref{eqn:loiconstgene2}). For instance, the density of contacts
and some parametrization of
the distribution of their orientation (called \emph{fabric} or \emph{texture}) have been studied, their
evolution can be related to strains~\cite{CCL97,RR01} (eqn.~\ref{eqn:loiconstgene2})
and their values can be correlated to the possibly
supported stress orientations~\cite{TRRC01} (role of $\alpha$ in eqn.~\ref{eqn:loiconstgene}). 
A brief presentation of the possible use of packing fraction
and fabric as work-hardening variables in a plasticity theory is ref.~\cite{RR98}.

Finally, let us briefly speculate about possible consequences of the existence
of strictly quasi-static and fragile regimes.
Sand specimens were observed to creep: under constant stresses~\cite{dBT97,dPI97} (e.g., on
stopping a triaxial test and maintaining constant stresses) strains vary
very slowly, over hours. (Let us quote ref.~\cite{RO62} again: `The time to equilibrium increases
to many days as the peak strength is approached').
\begin{figure}
 \centering 
 \includegraphics[width=6.5cm]{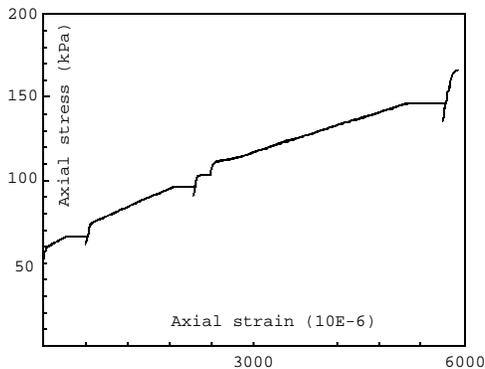}
\caption{Axial stress vs. axial strain curve, from ref.~\cite{dBT97}. 
The experiment was stopped at constant stress several times,
hence the creep intervals. Note the stiff response when the constant rate compression is resumed.
\label{fig:fluage}
}
\end{figure}
A new equilibrium might be approached, which can be rather
distant. When the slow controlled strain rate is resumed, the response to both positive and
negative $q$ increments is quite stiff. With due account to possible
aging phenomena in the contacts~\cite{elisabeth}, it is tempting to suggest the following explanation:
once left to wait at constant stress,
the grain pack, which is highly sensitive to noise, slowly drifts in configuration space, until it reaches
a state with a finite stability range (hence a stiff response of the `strictly quasi-static' type).
There appears thus to be interesting connections between slow dynamics and fragility.


\begin{thebibliography}{10}
\bibitem{GDV84}
{\em Constitutive Relations for Soils}, edited by G. Gudehus, F. Darve, and I.
  Vardoulakis (Balkema, Rotterdam, 1984).

\bibitem{Darve87}
{\em Manuel de rh\'eologie des g\'eomat\'eriaux}, edited by F. Darve (Presses
  des Ponts et Chauss\'ees, Paris, 1987).

\bibitem{DMWood}
Wood D.~M., {\em Soil Behaviour and Critical State Soil Mechanics} (Cambridge
  University Press, 1990).

\bibitem{KI01}
{\em Powders and Grains 2001}, edited by Y. Kishino (Balkema, Lisse, 2001).

\bibitem{HHL98}
{\em Physics of Dry Granular Media}, edited by H.~J. Herrmann, J.~P. Hovi, and
  S. Luding (Balkema, Dordrecht, 1998).

\bibitem{BH94}
Biarez, J. and Hicher, P.-Y., {\em Elementary Mechanics of Soil Behaviour}
(Balkema, Rotterdam, 1994)

\bibitem{CJ01}
Cambou B. and Jean M., {\em Microm\'ecanique des mat\'eriaux granulaires}
  (Herm\`es, Paris, 2001).

\bibitem{Ragneau}
Ragneau E., Ph.D. thesis, Institut National des Sciences Appliqu\'ees, Rennes,
  1993.

\bibitem{Boufellouh}
Boufellouh S., Ph.D. thesis, Ecole Centrale, Ch\^atenay-Malabry, 2000.

\bibitem{CWBC98}
Cates M.~E., Wittmer J.~P., Bouchaud  J.-P., and Claudin P., Phil. Trans. 
  Roy. Soc. London {\bf 356},  2535--2560  (1998).

\bibitem{dBGSC99}
Di~Benedetto H., Geoffroy H., Sauz\'eat C., and Cazacliu B.,  in {\em
  Pre-failure Deformation Characteristics of Geomaterials}, edited by M.
  Jamiolkowski, R. Lancellotta, and D. Lo~Presti (Balkema, Rotterdam, 1999),
  pp.\ 89--96.

\bibitem{HI96}
Hicher P.-Y., ASCE J. Geotechn. Eng., {\bf 122}, 641--648 (1996)

\bibitem{LA87}
Lanier J.,  in ref.\ \cite{Darve87}, pp.\ 15--31.

\bibitem{DCMM96}
Desrues J., Chambon R., Mokni M., and Mazerolle F., G\'eotechnique {\bf 46},
  529--546  (1996).

\bibitem{SAL77}
Salen\c{c}on J., {\em Applications of the Theory of Plasticity in Soil
  Mechanics} (Wiley, Chichester, 1977).

\bibitem{VER98}
Vermeer P.~A.,  in ref.\ \cite{HHL98}, pp.\ 163--196.

\bibitem{NO82}
Nova R., in ref.\ \cite{GDV84}, pp.\ 289--309.

\bibitem{WBK96}
Wu W., Bauer E., and Kolymbas D., Mech. of Materials {\bf 23},  45--69  (1996).

\bibitem{TY91}
Thornton C. and Yin K.~K., Powder Techn.{\bf 65},  153--166  (1991).

\bibitem{LJ00}
Lanier J. and Jean M., Powder Techn. {\bf 109},  206--221  (2000).

\bibitem{TH00}
Thornton C., G\'eotechnique {\bf 50},  43--53  (2000).

\bibitem{JNR2000}
Roux J.-N., Phys. Rev. E {\bf 61},  6802--6836  (2000).

\bibitem{CR2000}
Combe G. and Roux J.-N., Phys. Rev. Lett. {\bf 85},  3628--3631  (2000).

\bibitem{Gael}
Combe G., Ph.D. thesis, Ecole Nationale des Ponts et Chauss\'ees,
  Champs-sur-Marne (France) 2001.

\bibitem{CWBC98b}
Cates M.~E., Wittmer J.~P., Bouchaud J.-P. , and Claudin P., Phys. Rev. Lett.
  {\bf 81},  1841--1844  (1998).

\bibitem{BG90}
Bouchaud J.-P. and Georges A., Physics Reports {\bf 195},  127  (1990).

\bibitem{JNR97b}
 Roux J.-N.,  in {\em Proceedings of the Saint-Venant Symposium on Multiple
  Scale Analysis and Coupled Physical Systems} (Presses de l'Ecole Nationale
  des Ponts et Chauss\'ees, Paris, 1997), pp.\ 577--584.

\bibitem{CR01}
Combe G. and Roux J.-N.,  in ref.\ \cite{KI01}, pp.\ 293--296.

\bibitem{Narimane}
Bourada-Benyamina N., Ph.D. thesis, Ecole Nationale des Ponts et Chaussées,
  Champs-sur-Marne (France) 1999.

\bibitem{KAK01}
Kishino Y., Akaizawa H., and Kaneko K.,  in ref.\ \cite{KI01} pp.\ 199--202.

\bibitem{CRG90}
{\em Disorder and fracture}, edited by J.-C. Charmet, S. Roux, and E. Guyon
  (Plenum, New York, 1990).

\bibitem{WR97}
Williams J.~R. and Rege N., Powder Techn. {\bf 90},  187--194  (1997).

\bibitem{KU99}
Kuhn M.~R., Mech. of Materials {\bf 31},  407--429  (1999).

\bibitem{TKKR00}
T\"or\"ok J., Krishnamurthy S., Kert\'esz J., and Roux S.,
Phys. Rev. Lett. {\bf 84},  3851--3854  (2000).

\bibitem{RO62}
Rowe P.~W., Proc. Roy. Soc. London {\bf A269},  500--526 (1962).

\bibitem{CCL97}
Calvetti F., Combe G., and Lanier J., Mech. Coh.-Frict. Mat.
{\bf 2},  121--163  (1997).

\bibitem{RR01}
Radjai F. and Roux S.,  in ref.\ \cite{KI01} pp.\ 21--24.

\bibitem{TRRC01}
Troadec H., Radjai F., Roux S., and Charmet J.-C.,
in ref.\ \cite{KI01} pp.\ 25--28.

\bibitem{RR98}
Roux S. and Radjai F.,  in ref.\ \cite{HHL98}, pp.\ 229--235.

\bibitem{dBT97}
Di~Benedetto H. and Tatsuoka F., Soils and Foundations {\bf 37}, 127--138  (1997).

\bibitem{dPI97}
Di~Prisco C. and Imposimato S., Mech. Coh.-Frict. Mat. {\bf 2}, 93--120  (1997).

\bibitem{elisabeth}
Charlaix E., in this issue.
\end{thebibliography}
\end{document}